\documentclass[aps,prb,twocolumn,superscriptaddress]{revtex4}
\usepackage{tabularx,graphicx}
\usepackage{latexsym}
\usepackage{amsmath, amsthm, amssymb}
\usepackage{bbm}

\let\z=\zeta  \let\q=\theta

\def\nn{\nonumber}

\def\be{\begin{equation}}
\def\ee{\end{equation}}
\def\bea{\begin{eqnarray}}
\def\eea{\end{eqnarray}}
\def\ba{\begin{array}}
\def\ea{\end{array}}

\begin{document}

\title{Quantum criticality of the kagome antiferromagnet\\ with Dzyaloshinskii-Moriya interactions}

\author{Yejin Huh}
\affiliation{Department of Physics, Harvard University, Cambridge, Massachusetts 02138, USA}

\author{Lars Fritz}
\affiliation{Department of Physics, Harvard University, Cambridge, Massachusetts 02138, USA}

\affiliation{Institut f\"ur Theoretische Physik, Universit\"at zu K\"oln,
Z\"ulpicher Stra\ss e 77, 50937 K\"oln, Germany}

\author{Subir Sachdev}
\affiliation{Department of Physics, Harvard University, Cambridge, Massachusetts 02138, USA}

\date{\today}

\begin{abstract}
We investigate the zero-temperature phase diagram of the nearest-neighbor kagome antiferromagnet in the presence of Dzyaloshinksii-Moriya interaction. We develop a theory for the transition between $Z_2$ spin liquids with bosonic spinons
and a phase with antiferromagnetic long-range order. Connections to recent numerical studies
and experiments are discussed.
\end{abstract}
\pacs{}

\maketitle


\section{Introduction}\label{intro}

The nearest neighbor spin $S=\frac{1}{2}$ antiferromagnet on the kagome lattice 
has been the focus of extensive theoretical and experimental study because 
it is a prime candidate for realizing a ground state without antiferromagnetic order.

On the experimental side, much attention has focused on the $S=1/2$ compound herbertsmithite
ZnCu$_3$(OH)$_6$Cl$_2$. It has $J \approx 170$ K and 
no observed ordering or structural distortion. \cite{helton,ofer,mendels,imai}
However, there is an appreciable amount of substitutional disorder between the Zn and Cu sites (believed to be of the order 5-10 $\%$)
which affects the low $T$ behavior. \cite{bert,gregor,chitra,ka5,mila1}
More importantly, there is an upturn in the susceptibility at $T=75$ K which has been ascribed to the
DM interactions \cite{rigol,zorko,ofer2,mila1,mila2}.

On the theoretical side, the most recent evidence \cite{kag2,kag3,kag4,kag5,poilblanc1,poilblanc2} on the 
the nearest-neighbor antiferromagnet points consistently
to a ground state with a spin gap of $0.05J$ and valence bond solid (VBS) order. The pattern of the VBS order
is quite complex with a large unit cell, but was anticipated by Marston and Zeng \cite{kag1}
by an application of the VBS selection mechanism described in the $1/N$ expansion
of the SU($N$) antiferromagnet. \cite{rs1}

The influence of the DM interactions has also been studied theoretically\cite{cepas,shtengel,ran2,mila2}.
Starting with an ``algebraic spin liquid'' ground state, Hermele {\em et al.} \cite{ran2}
argued that the DM coupling, $D$,  was a relevant perturbation, implying that an infinitesimal $D$
would induce long-range magnetic order. 
In a recent exact diagonalization study, Cepas {\em et al.} \cite{cepas} 
reach a different conclusion: they claim that there is a non-zero critical DM coupling $D_c$ beyond
which magnetic order is induced. They estimate $D_c/J \approx 0.1$, quite close
to the value measured\cite{zorko} for ZnCu$_3$(OH)$_6$Cl$_2$ which has $D/J \approx 0.08$.
This proximity led Cepas {\em et al.} to suggest that the quantum criticality of the DM-induced
transition to magnetic order controls the observable properties of this kagome antiferromagent.

The purpose of this paper is to propose a theory for the quantum critical point discovered
by Cepas {\em et al.} \cite{cepas}. We will compute various observables of this theory, allowing
a potential comparison with numerics and experiments. 

Given the evidence for VBS order
in the model without DM interactions\cite{kag2,kag3,kag4,kag5}, it would appear we need
a theory for the transition from the VBS state to the magnetically ordered state. However, the VBS ordering
is weak, and can reasonably be viewed as a perturbation on some underlying spin-liquid ground state.
Schwandt, Mambrini, and Poilblanc \cite{poilblanc1,poilblanc2} have recently presented evidence that
the kagom\'e antiferromagnet is proximate to a $Z_2$ spin liquid state, and that vison condensation in this state
leads to weak VBS ordering. Their dimer representation leads naturally to $Z_2$ spin liquid
states in the same
class as that originally described \cite{self6,wang} by the Schwinger boson method \cite{assa,leshouches}.
We will therefore neglect the complexities associated with the VBS ordering and work
with the parent $Z_2$ spin liquid state. 
This is equivalent to ignoring the physics of the vison sector,
and assumes that the magnetic ordering transition can be described in a theory of the spinons alone \cite{cenke}. 
The main result of this paper will be a theory
of the quantum phase transition from the Schwinger boson $Z_2$ spin liquid
to the magnetically ordered state as induced by the DM interactions.

We will begin in Section~\ref{sec:model} with a description of the mean-field theory 
of the $Z_2$ spin liquid and its transition to the magnetically ordered state 
in the presence of DM interactions. This will be carried using the
Sp($N$) Schwinger boson formulation \cite{self6,leshouches}, for which the mean-field theory becomes 
exact in the large $N$ limit. We will turn to fluctuation corrections and the nature of 
the quantum critical point in Section~\ref{sec:critical}. Here we will show
that the critical theory is the familiar three dimensional XY model. However, its connection to 
experimental observables is subtle: in particular, the XY field itself is not directly observable.

While this paper was in preparation, a description of the Schwinger boson mean field theory
in the presence of DM interactions also appeared in Ref.~\onlinecite{cepas2}; they consider mean-field
solutions with larger unit cells than we do, but did not analyze the critical field theory.
Where they overlap,
our results are in agreement with theirs. We also note the recent experimental observations of Helton
{\em et al.} \cite{helton2}, who present evidence for quantum criticality in ZnCu$_3$(OH)$_6$Cl$_2$.

\section{Mean field theory}
\label{sec:model}

\begin{figure}[t]
\includegraphics[width=0.45\textwidth]{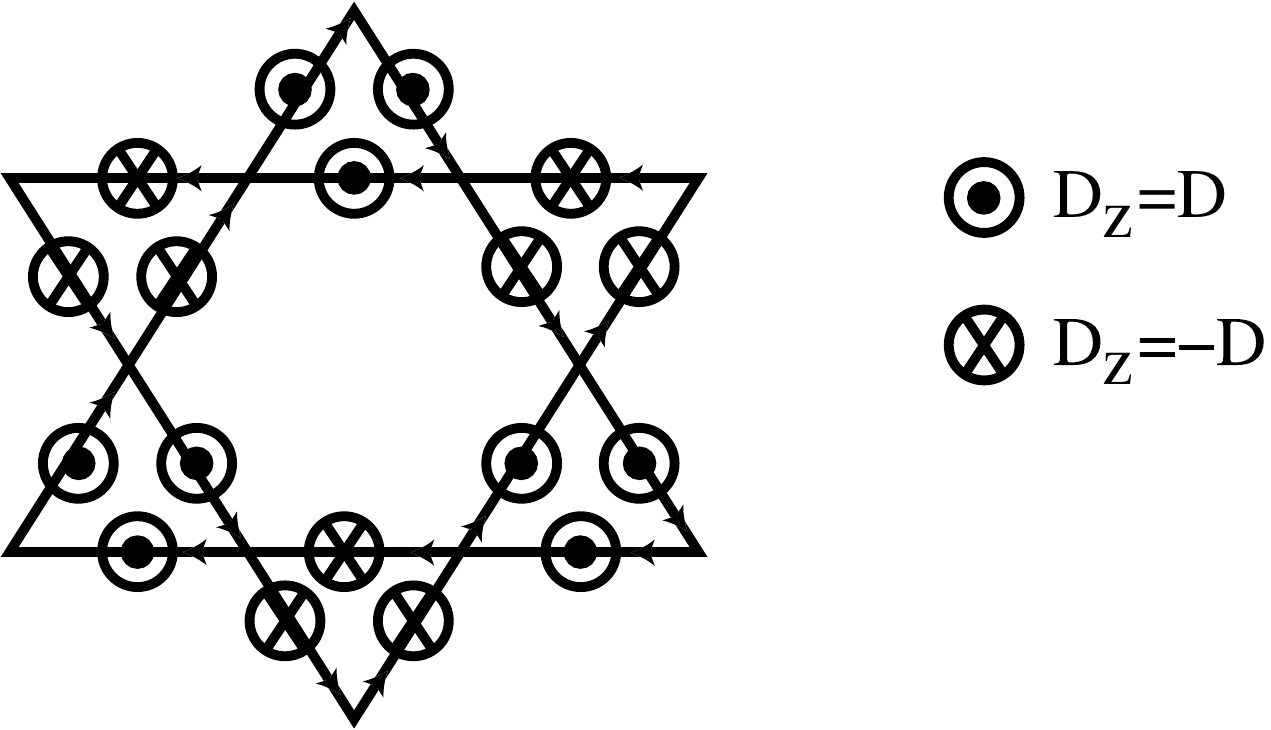}
\caption{Staggered DM interaction from triangle to triangle in $z$-direction as in Ref.~\onlinecite{Elhajal}. The arrowheads indicate $D$ to come out of the plane, whereas the tails denote $D$ to go into the plane. Note that the triangles are summed clock- and anticlockwise, respectively, indicated by the arrows on the bonds.}\label{fig:DM}
\end{figure}

The model we consider is a standard Heisenberg Hamiltonian supplemented by an additional DM interaction. It assumes the form
\begin{eqnarray}
\mathcal{H}=\frac{1}{2}\sum_{i,j} \left[J_{ij} \bf{S}_i \cdot \bf{S}_j + \bf{D}_{ij}\cdot \left (\bf{S}_i \times \bf{S}_j \right) \right]\;.
\end{eqnarray}
$\bf{S}_i$ in this notation denotes the spin operator at site $i$, $J_{ij}$ is assumed to be uniform and of the nearest neighbor type, and ${\bf{D}}_{ij}=D_{ij}{\bf{e}}_z$ is taken along the z-axis and staggered from triangle to triangle~\cite{Elhajal}, see Fig.~\ref{fig:DM}. 
This additional term explicitly breaks the spin-rotation symmetry by favoring configurations lying in the x-y plane. Furthermore this term increases the tendency of classical spin ordering. It has been shown in earlier works~\cite{assa,leshouches} that Schwinger bosons are ideally suited to describe phase transitions between paramagnetic and magnetically ordered phases in spin models. Following Ref.~\onlinecite{self6} we introduce a Sp($N$) 
generalization of the spin operators, which formally allows to consider a controlled large $N$ limit particularly suited for the study of frustrated spin systems such as kagome or traingular antiferromagnets. In the Sp(1) case, which is isomorphic to SU(2), one can represent the spin variables as
\begin{eqnarray}
{\bf{S}}_i={b}^\ast_{i\sigma} \frac{{\boldsymbol \tau}_{\sigma\sigma'}}{2}{{b}}^{\phantom{\ast}}_{i\sigma'}
\end{eqnarray}
with ${\vec{\tau}}$ being the Pauli matrices and with
\begin{eqnarray}\label{Eq:constraint}
{{b}}^{\phantom{\ast}}_{i\sigma}=\left ( \begin{array} {c} b^{\phantom{\ast}}_{i\uparrow} \\ b^{\ast}_{i\downarrow} \end{array}\right) \quad {\rm{and}} \quad \sum_\sigma {{b}}^\ast_{i\sigma} {{b}}^{\phantom{\ast}}_{i\sigma}=2S=n_b\;.
\end{eqnarray}
In the case of the large-$N$ generalization the Schwinger bosons acquire another index counting the copies of the system (we drop this index in the following discussions, but display $N$ whenever it is essential). The large-$N$ 
generalization formally justifies the mean-field, with the saddle point becoming exact in the limit $N\to \infty$. In order to properly reformulate the problem at hand we introduce two decoupling parameters
\begin{eqnarray}\label{Eq:HS}
Q_{ij} &=& \sum_{\sigma \sigma'} \epsilon_{\sigma \sigma'} b_{i\sigma} b_{j\sigma'} \nonumber \\ 
P_{ij} &=& \sum_{\sigma \sigma'} \tau^x_{\sigma \sigma'} b_{i\sigma} b_{j\sigma'} 
\end{eqnarray}
where $\epsilon_{\sigma \sigma'}$ is the antisymmetric tensor and $\tau^x$ is just the standard Pauli matrix. We see from the above expressions that $Q_{ij}=-Q_{ji}$ whereas $P_{ij}=P_{ji}$. This implies that the bond variables $P_{ij}$ do not have a direction.

The constraint Eq.~\eqref{Eq:constraint} is implemented via a Lagrangian multiplier in a standard way. The Hamiltonian of the system formulated in the fields defined in Eq.~\eqref{Eq:HS} consequently reads
\begin{eqnarray}
\frac{\mathcal{H}}{N}=&-&\frac{1}{2}\sum_{i,j} J_{ij}Q_{ij}^\ast Q^{\phantom{\ast}}_{ij}-\frac{i}{4}\sum_{i,j} D_{ij} (P_{ij}^\ast Q^{\phantom{\ast}}_{ij}-Q_{ij}^\ast P^{\phantom{\ast}}_{ij})\nonumber \\ &+& \sum_i \lambda_i \left(b^\ast_{i\sigma}b^{\phantom{\ast}}_{i\sigma}-\kappa \right)
\label{eqn:H exp}
\end{eqnarray}
where $\kappa={n_b}/{N}$. We furthermore introduce $N_u$ as the number of unit cells in the systems and $N_s$ as the number of sites within the unit cell. We can write the mean-field Hamiltonian per flavor and unit cell as
\begin{eqnarray}\label{eq:HMF}
\frac{H_{MF}}{N N_u} &=& \frac{1}{N_u}\sum_{\bf{k}} \Psi^\ast ({\bf{k}})\mathbb{H}(q_{ij},p_{ij},\lambda,{\bf{k}}) \Psi({\bf{k}})\nonumber \\ &+&\frac{J}{2}\sum_{(ij)'}|q_{ij}|^2+\frac{iD}{4}\sum_{(ij)'}(p_{ij}^*q_{ij}-q_{ij}^*p_{ij})\nonumber \\ &-&N_s\lambda(1+\kappa)
\end{eqnarray}
where $\sum_{(ij)'}$ denotes the sum over bonds belonging to the unit cell,
\begin{eqnarray}\label{eq:Psi}
\Psi^\ast ({\bf{k}})=(b^\ast_{1\uparrow}({\bf{k}}),...,b^\ast_{N_s\uparrow}({\bf{k}}),b^{\phantom{\ast}}_{1\downarrow}(-{\bf{k}}),...,b^{\phantom{\ast}}_{N_s\downarrow}(-{\bf{k}}))  \;,
\end{eqnarray}
and the matrix
\begin{eqnarray}
\mathbb{H}=\left( \begin{array} {cc} \lambda \mathbb{I} & \mathbb{C}^\ast(k) \\ \mathbb{C}(k) & \lambda \mathbb{I} \end{array} \right) 
\end{eqnarray}
with the matrices $\mathbb{I}$ (identity) and $\mathbb{C}(k) $ being $N_s\times N_s$ matrices; the explicit
form of these matrices is given in Appendix~\ref{sec:micro}. 
As mentioned before, one of the major assets of the Schwinger boson approach is that it can describe magnetically disordered gapped spin liquid phases as well as magnetically ordered states. On a formal level in the large $N$ 
approach this is achieved by introducing the following notation for the Schwinger bosons 
\begin{eqnarray}
{b}_{i\sigma} =(\sqrt{N}x_i,b_i^{\tilde{m}}) \quad {\rm{where}} \quad \tilde{m}=2,...,N \;.
\end{eqnarray}
The first component is thus a classical field. If $x\neq0$ it signals condensation which causes long range order to appear.
Following Ref.~\onlinecite{assa,self6} we can integrate out the Schwinger bosons and the zero-temperature mean field energy assumes the following form
\begin{eqnarray}
\frac{E_{MF}}{N N_u} &=& \frac{1}{N_u}\sum_{{\bf{k}},\mu=1,..,N_s} \omega_\mu({\bf{k}})-N_s\lambda(1+\kappa)+\lambda\sum_{i'\sigma} |x_{i\sigma}|^2\nonumber \\ &+&\frac{J}{2}\sum_{(ij)'}\left[ |q_{ij}|^2-\left(q^*_{ij}\epsilon_{\sigma \sigma'}x_{i\sigma}x_{j\sigma'})+{\rm{h.c.}}\right) \right]\nonumber \\ &+&\frac{iD}{4}\sum_{(ij)'}(p_{ij}^*q_{ij}-q_{ij}^*p_{ij})\nonumber \\ &-& \frac{iD}{4}\sum_{(ij)'}(p_{ij}^*\epsilon_{\sigma \sigma'}x_{i\sigma}x_{j\sigma'}+q_{ij}\tau^x_{\sigma \sigma'}x^*_{i\sigma}x^*_{j\sigma'})\nonumber \\&+& \frac{iD}{4}\sum_{(ij)'}(p_{ij}\epsilon_{\sigma \sigma'}x^*_{i\sigma}x^*_{j\sigma'}+q_{ij}^*\tau^x_{\sigma \sigma'}x_{i\sigma}x_{j\sigma'})
\end{eqnarray}
where $\sum_{i'}$ denotes the sum over all sites within one unit cell.
In the following we solve the self-consistency equations according to
\begin{eqnarray}\label{Eq:MF}
\kappa&=& \langle {{b}}^{\ast}_{i\sigma} {{b}}^{\phantom{\ast}}_{i\sigma} \rangle_{MF}  \nonumber  \\q_{ij}&=&\langle Q_{ij}\rangle_{MF}\;, \nonumber  \\ p_{ij}&=&\langle P_{ij}\rangle_{MF} \;.
\end{eqnarray}
with the Hamiltonian defined in Eq.~\eqref{eq:HMF}.

\begin{figure}
\includegraphics[width=0.45\textwidth]{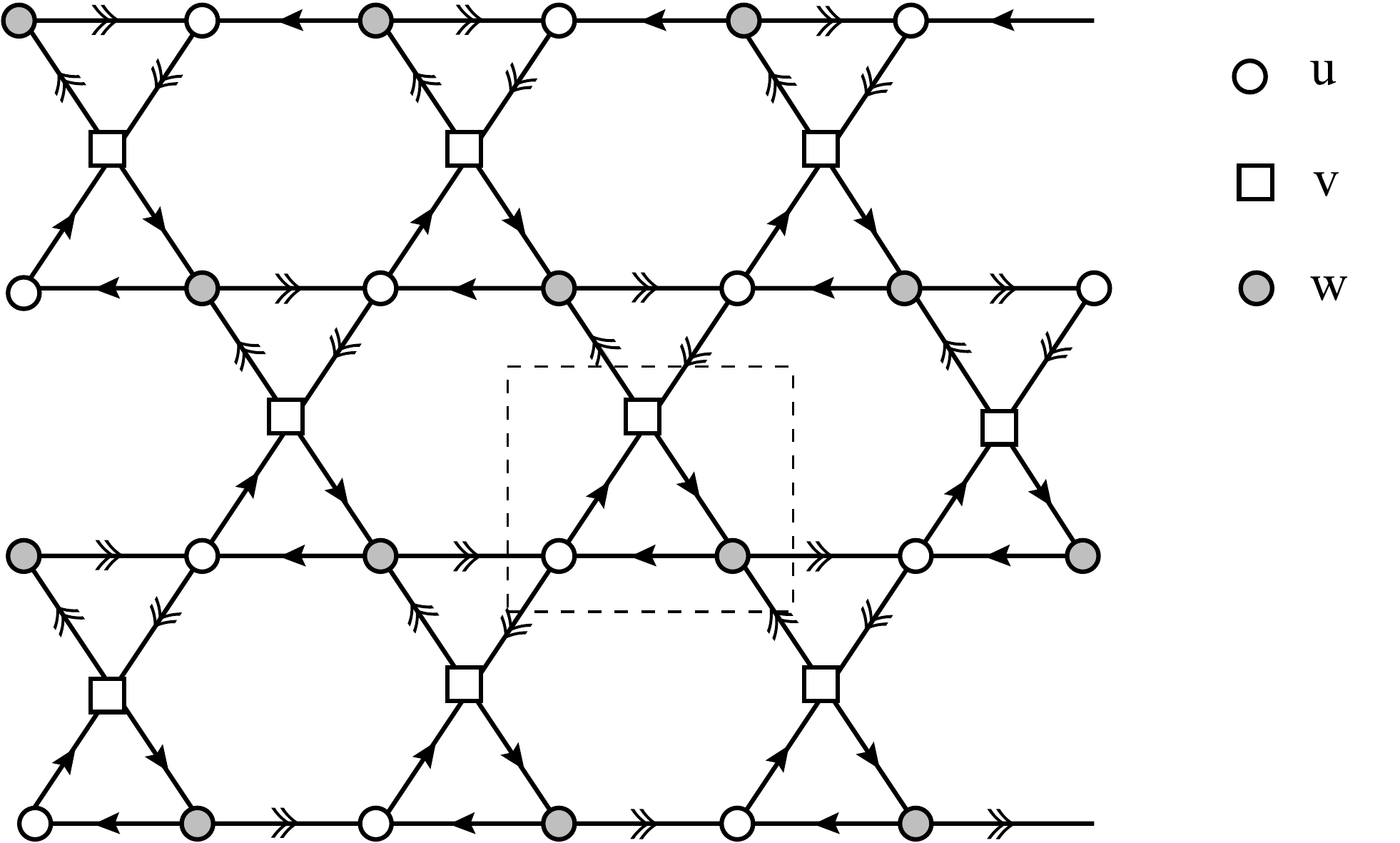}
\caption{We take a unit cell of three sites, labeled $u$, $v$, $w$. The arrows indicate the values of the oriented variables $q_{ij}$. The links with single arrows have $q_{ij} = q_1$, while those with double arrows have $q_{ij} = q_2$.
The $P_{ij}$ are unoriented, and take values $p_1$ and $p_2$ on these links respectively.}\label{fig:dimercover}
\end{figure}
Our solution of the mean-field equations follows previous work \cite{self6,wang}, which classified physically
different $Z_2$ spin liquid solutions without the DM interactions. We found that these solutions have
a natural generalization in the presence of DM terms, with values of the $p_{ij}$ which reflect the
symmetries of the $q_{ij}$. Two stable solutions were found in previous work, with only
two possibly distinct values of $q_{ij}$ as illustrated in Fig.~\ref{fig:dimercover}. Including the DM interactions,
these solutions extended to\\
({\em i\/}) $q_1=q_2$ real, $p_1=p_2$ pure imaginary: upon increasing $\kappa$, the Schwinger bosons
condense at ${\bf k}=0$, with the spins at angles of 120$^\circ$ to each other within the unit cell.
This states is therefore called the ${\bf k}=0$ state.\\
({\em ii\/}) $q_1=-q_2$ real, $p_1=-p_2$ pure imaginary: upon increasing $\kappa$, the Schwinger bosons
condense at wavevector ${{\bf k}}=\pm(2\pi/3,0)$ into a state which is called 
the $\sqrt{3}\times\sqrt{3}$ antiferromagnet, characterized by an enlarged unit cell.\\
Solutions with larger unit cells can be present with additional frustrating interactions \cite{wang}, but we
will not consider them here.

\subsection{Phase diagram}
\label{sec:phasediagram}

\begin{figure}[h]\label{fig:phasediagram} 
\includegraphics[width=0.5\textwidth]{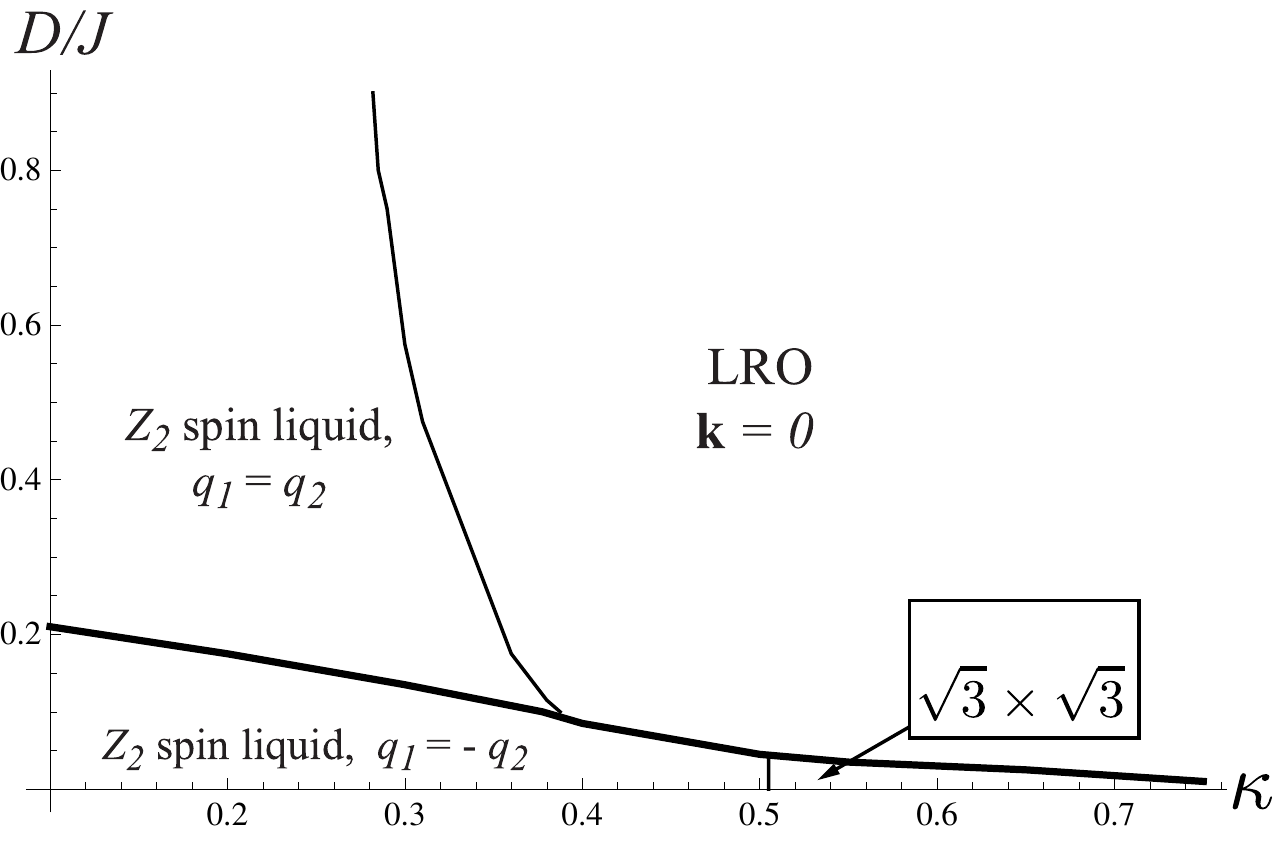}
\caption{Phase diagram in the $N\rightarrow \infty$ limit of the Sp($N$) theory. 
The x-axis shows $\kappa={2S}/{N}$ and the y-axis $D/J$. The phases have long-range magnetic order type (LRO), or gapped $Z_2$ spin liquids. The thick line is a first-order transition, while the thin lines are second-order transitions:
the transition to the ${\bf k} =0$ LRO state is described in Section~\ref{sec:critical}. The limit of $D/J\to 0$ reduces to the results presented in Ref.~\onlinecite{self6}.}
\end{figure}

Our phase diagram is shown in Fig.~\ref{fig:phasediagram}
as a function of $\kappa = n_b/N$ (which corresponds to the spin size) and the parameter $D/J$. 
Our phase diagram is similar to that obtained recently by Messio {\em et al.}~\cite{cepas2}. 
They also considered solutions with a larger unit cell which were stable over some portion of the phase diagram.

We start with a discussion of the classical limit. While for $D=0$ the long-range ordered state of the $\sqrt{3}\times \sqrt{3}$ type is generically preferred~\cite{self6}, infinitesimal $D$ favors the so-called $\mathbf{k}=0$ state. 
This is reproduced by our mean-field equations in the large spin limit $\kappa\to \infty$. 
For finite values of $\kappa$ there is a small slab in which long-range order of the $\sqrt{3}\times \sqrt{3}$ type is favored over the $\mathbf{k}=0$. These states are separated by a first order transition driven by the ratio $D/J$. 

A similar behavior appears for the corresponding spin liquid states at small $\kappa$. The $q_1=-q_2$
is favored at small $D/J$, and then undergoes a first order transition to the $q_1 = q_2$ state at large $D/J$.

In exact diagonalization studies of the spin $S=\frac{1}{2}$ kagome antiferromagnet with DM interactions a second order phase transition between a phase with short ranged $\mathbf{k}=0$ correlations~\cite{kag4,Laeuchli} (obtained in the pure Heisenberg case) and phase with $\mathbf{k}=0$ long range order was found~\cite{cepas2}. Such a transition is also present
in our mean-field theory, which therefore can be used for a study of critical properties in Section~\ref{sec:critical}.

\section{Quantum criticality}
\label{sec:critical}

We will consider only the transition out of the $q_1=q_2$ spin liquid, because that is what is seen in the numerical
studies \cite{cepas}. The corresponding transition out of the $q_1 = - q_2$ spin liquid can be treated in a similar manner.
Throughout this section we will consider the physical SU(2) antiferromagnet directly, and not take the large $N$ limit.
The method followed below has been reviewed in a more general context in Ref.~\onlinecite{leshouches}.

Since we are at ${\bf k} =0$, we can write the effective action for the bosons by making the small momentum
expansion of the matrix in Eq.~(\ref{eq:cmat}). We take 3 sites, $u$, $v$, $w$, in each unit cell (see Fig.~\ref{fig:dimercover}),
and then take the continuum limit of the saddle-point Lagrangian. We write the boson operators on these
sites as $b_{u\sigma} = U_\sigma$, $b_{v \sigma} = V_\sigma$ etc.,  and set $q_1 = q_2 = q$, and $p_1=p_2 = i p$ 
with $q$ and $p$ real.
Then, we write the final Lagrangian in the form
\begin{equation}
\mathcal{L} = \mathcal{L}_H + \mathcal{L}_{DM} 
\end{equation}
representing the contributions of the Heisenberg exchange and the DM coupling, respectively. 

From Eq.~(\ref{eq:cmat}) we obtain the Lagrangian in the absence of a DM term (which describes the ${\bf k=0}$
solution of Ref.~\onlinecite{self6}):
\begin{eqnarray}\label{eq:lagra}
\mathcal{L}_H &=& U^\ast_\sigma \frac{\partial U_\sigma}{\partial \tau} + V^\ast_\sigma \frac{\partial V_\sigma}{\partial \tau} + W^\ast_\sigma \frac{\partial W_\sigma}{\partial \tau} \nonumber \\
&~&~~+ \lambda \left( |U_\sigma|^2 + |V_\sigma|^2 + |W_\sigma|^2
\right) \nonumber \\ &~&~~- J q \epsilon_{\sigma\sigma'} \left( U_\sigma V_{\sigma'} + V_\sigma W_{\sigma'} 
+ W_\sigma U_{\sigma'} \right) + \mbox{c.c.}  \nonumber \\ &+&
\frac{J q}{2} \epsilon_{\sigma\sigma'} \left(  \partial_1 U_\sigma \partial_1 V_{\sigma'} + \partial_2 V_\sigma 
\partial_2 W_{\sigma'} + \partial_3 W_\sigma  \partial_3 U_{\sigma'} \right) \nonumber \\
&~&~~+ \mbox{c.c.}
\end{eqnarray}
where $\partial_i$ is the gradient along the direction ${\bf e}_i$ in Eq.~(\ref{eq:direct}).

We now perform a unitary transformation to new variables $X_\sigma$, $Y_\sigma$, $Z_\sigma$. These
are chosen to diagonalize only the non-gradient terms in $\mathcal{L}_H$. 
\begin{eqnarray}
 \left( \begin{array}{c} U_\sigma \\ V_\sigma \\ W_\sigma  \end{array} \right) &=& \frac{Z_\sigma}{\sqrt{6}} \left( \begin{array}{c}
1 \\ \zeta \\ \zeta^2  \end{array} \right)
 + \epsilon_{\sigma\sigma'}\frac{Z_{\sigma'}^\ast}{\sqrt{6}}  \left( \begin{array}{c}
i \\ i \zeta^2 \\  i \zeta  \end{array} \right) \nonumber \\
&+& \frac{Y_\sigma}{\sqrt{6}} \left( \begin{array}{c}
1 \\ \zeta \\ \zeta^2 \end{array} \right)
 + \epsilon_{\sigma\sigma'}\frac{Y_{\sigma'}^\ast}{\sqrt{6}}  \left( \begin{array}{c}
-i \\ -i \zeta^2 \\ - i \zeta  \end{array} \right) \nonumber \\
&~&~~~~~+ \frac{X_\sigma}{\sqrt{3}} \left( \begin{array}{c} 1 \\ 1 \\ 1 \end{array} \right) .
 \label{umat}
\end{eqnarray}
where $\zeta \equiv e^{2 \pi i /3}$. The tensor structure above makes it clear
that this transformation is rotationally invariant, and that $X_\sigma$, $Y_\sigma$, $Z_\sigma$ 
transform as spinors under SU(2) spin rotations. Inserting Eq.~(\ref{umat}) into $\mathcal{L}_H$ we find
\begin{eqnarray}
\mathcal{L}_H &=& X_\sigma^\ast \frac{\partial X_\sigma}{\partial \tau} + Y_\sigma^\ast \frac{\partial Z_\sigma}{\partial \tau} + Z_\sigma^\ast \frac{\partial Y_\sigma}{\partial \tau} + 
 (\lambda + \sqrt{3} Jq) |Z_\sigma |^2 \nonumber \\
 &~&~~~~+ (\lambda - \sqrt{3} Jq) |Y_\sigma|^2 + 
\lambda |X_\sigma |^2 \nonumber \\
&~&~~~~~~+ \frac{J q \sqrt{3}}{2} \left( |\partial_x Z_\sigma |^2 + |\partial_y Z_\sigma |^2 \right) + \ldots
\label{lz1}
\end{eqnarray}
The ellipses indicate omitted terms involving 
spatial gradients in the $X_\sigma$ and $Y_\sigma$ which we will not keep track of.
This is because the fields $Y_\sigma$ and $X_\sigma$ are massive relative to $Z_\sigma$ (for $q<0$ which is the case in our mean-field solution), 
and so can be integrated out. This yields the effective Lagrangian
\begin{eqnarray}
\mathcal{L}_H^Z &=& \frac{1}{(\lambda - \sqrt{3}J q) } |\partial_\tau Z_\sigma |^2 + 
\frac{J q \sqrt{3}}{2} \left( |\partial_x Z_\sigma |^2 + |\partial_y Z_\sigma |^2 \right) \nonumber \\
&~&~~~~+ (\lambda + \sqrt{3} Jq) |Z_\sigma |^2 + \ldots \label{lhz}
\end{eqnarray}
Note that the omitted spatial gradient terms in $X_\sigma$, $Y_\sigma$ do contribute a correction to the
spatial gradient term in Eq.~(\ref{lhz}), and we have not accounted for this. This
Lagrangian shows that the mean-field theory has a transition to magnetic order
at $\lambda = |\sqrt{3}J q|$, which agrees with earlier results \cite{self6}. 

The effective Lagrangian $\mathcal{L}_H^Z$ is almost the complete solution for the critical theory 
in the system without the DM interactions. However, we also need higher
order terms in Eq.~(\ref{lhz}), which will arise from including the fluctuations of the gapped fields
$Q$ and $\lambda$. Rather than computing these from the microscopic Lagrangian, it is more efficient
to deduce their structure from symmetry considerations. The representation in Eq.~(\ref{umat}), and the 
connection of the $U$, $V$, $W$ to the lattice degrees of freedom, allow us to deduce the
following symmetry transformations of the $X$, $Y$, $Z$:
\begin{itemize}
\item Under a global spin rotation by the SU(2) matrix $g_{\sigma\sigma'}$, we have $Z_\sigma 
\rightarrow g_{\sigma\sigma'} Z_{\sigma'}$, and similarly for $Y$, and $Z$.
When DM interactions are included, the global symmetry is reduced to
U(1) rotations about the $z$ axis, under which 
\begin{eqnarray}
Z_\uparrow \rightarrow e^{i \theta} Z_{\uparrow}~~&,&~~
Z_\downarrow \rightarrow e^{-i \theta} Z_{\downarrow} \nonumber \\
Y_\uparrow \rightarrow e^{i \theta} Y_{\uparrow}~~&,&~~
Y_\downarrow \rightarrow e^{-i \theta} Y_{\downarrow} \nonumber \\
X_\uparrow \rightarrow e^{i \theta} X_{\uparrow}~~&,&~~
X_\downarrow \rightarrow e^{-i \theta} X_{\downarrow}. \label{srot}
\end{eqnarray}
\item Under a 120$^\circ$ lattice rotation, we have $U_\sigma \rightarrow V_\sigma$, 
$V_\sigma \rightarrow W_\sigma$, $W_\sigma \rightarrow U_\sigma$. From (\ref{umat}), we see
that this symmetry is realized by 
\begin{equation}
Z_\sigma \rightarrow \zeta Z_\sigma~,~Y_\sigma \rightarrow \zeta Y_\sigma~,~
X_\sigma \rightarrow X_\sigma . \label{trot}
\end{equation}
Note that this is distinct
from the SU(2) rotation because $\mbox{det} (\zeta) \neq 1$.
\item
Under time-reversal, we have $U_\sigma \rightarrow \epsilon_{\sigma\sigma'} U^{\ast}_{\sigma'}$, and similarly
for $V_\sigma$, $W_\sigma$. This is realized in Eq.~(\ref{umat}) by 
\begin{equation}
Z_\sigma \rightarrow i Z_\sigma~,~
Y_\sigma \rightarrow - i Y_\sigma~,~ X_\sigma \rightarrow   \epsilon_{\sigma\sigma'} X^{\ast}_{\sigma'}. \label{trev}
\end{equation}
In particular, note that $Z_\uparrow$ does not map to $Z_\downarrow$ under time-reversal.
\end{itemize}
It is easy to verify that Eq.~(\ref{lz1}) is invariant under all the symmetry operations above.
These symmetry operators make it clear that the only allowed quartic term for the Heisenberg Hamiltonian
is $\left( \sum_\sigma |Z_\sigma|^2 \right)^2$: this implies that the $Z_2$ spin liquid to antiferromagnetic order
transition of this model is in the universality class of the O(4) model \cite{css}. 

Let us now include the DM interactions. From Eq.~(\ref{eq:cmat}), we see that
\begin{equation}
\mathcal{L}_{DM} =  i \frac{D q}{2} \tau^x_{\sigma \sigma'} \left( U_\sigma V_{\sigma'} + V_\sigma W_{\sigma'} + W_\sigma U_{\sigma'} \right) + \mbox{c.c.}.
\end{equation}
We have dropped a term proportional to $Dp$ which has the same structure as the terms in Eq.~(\ref{lz1}),
and ignored spatial gradients.
In terms of the fields in Eq.~(\ref{umat}), this takes the simple form
\begin{eqnarray} 
\mathcal{L}_{DM} &=& \frac{Dq}{2} \Bigl( i \tau^x_{\sigma\sigma'} X_\sigma X_{\sigma'} 
+ \mbox{c.c.} \nonumber \\
&~&~~~~~~~~~~- Z^\ast_{\sigma} \tau^z_{\sigma\sigma'} Z_{\sigma'} + Y^\ast_{\sigma} \tau^z_{\sigma\sigma'} Y_{\sigma'} \Bigr), \label{lz2} 
\end{eqnarray}
and it can be verified that these terms are invariant under Eqs.~(\ref{srot},\ref{trot},\ref{trev}).
As before, we now integrate out $X_\sigma$ and $Y_\sigma$ from $\mathcal{L}_H + \mathcal{L}_{DM}$ in  Eqs.~(\ref{lz1}) and (\ref{lz2}). 
We obtain a Lagrangian with the same structure as Eq.~(\ref{lhz}), but all couplings become dependent upon
$\sigma$; in other words, we have 2 separate XY models for $Z_\uparrow$ and $Z_\downarrow$.
Performing a careful analysis of allowed higher order terms as restricted by the symmetry constraints
discussed above, and after appropriate rescalings of the spatial, temporal, and field scales, we obtain the
field theory with the Lagrangian
\begin{eqnarray}
\mathcal{L}_Z &=& |\partial_\tau Z_\uparrow|^2 + |{\boldsymbol \nabla} Z_\uparrow |^2
+ s_\uparrow |Z_\uparrow |^2 + u_\uparrow |Z_\uparrow|^4 \nonumber \\
&+& |\partial_\tau Z_\downarrow|^2 + |{\boldsymbol \nabla} Z_\downarrow |^2
+ s_\downarrow |Z_\downarrow |^2 + u_\downarrow |Z_\downarrow|^4 \nonumber \\
&+& v |Z_\uparrow|^2 |Z_\downarrow|^2  + w \left( (Z_\uparrow Z_\downarrow)^6  +  (Z_\uparrow^\ast Z_\downarrow^\ast )^6 \right).
\label{lz}
\end{eqnarray}
Note that $s_\uparrow \neq s_\downarrow$ in general (and similarly for $u_{\uparrow,\downarrow}$ etc.), and equality
is not required by the time-reversal symmetry in Eq~(\ref{trev}). Time-reversal
symmetry does prohibit a term $\sim (Z_\uparrow Z_\downarrow)^3$ which is allowed by the other symmetries.
Thus we expect only one of $Z_\uparrow$ or $Z_\downarrow$ to condense at the quantum
critical point: as we will see from the analysis of observables in Section~\ref{sec:obs}, this transition does
indeed correspond to the development of spiral magnetic order in the $x$-$y$ plane. The choice between
$Z_\uparrow$ and $Z_\downarrow$ is controlled by the sign of $D$. 

Eq.~(\ref{lz}) also contains terms which couple the two XY models to each other. 
The lowest allowed term, $v$, couples the energy densities
and does not have any important effects. More interesting is the $w$ term,
which shows that the global symmetry is not O(2)$\otimes$O(2) but O(2)$\otimes \mathbb{Z}_{12}$.
In the magnetically ordered phase with $\langle Z_\uparrow \rangle \neq 0$ (say), this term will induce a small 
ordering field $\sim Z_\downarrow^6$ in the XY model for $Z_\downarrow$. However, the action for $Z_\downarrow$
has a `mass' term $s_\downarrow$ with a positive co-efficient, and this sixth order term will not immediately
induce ordering in $Z_\downarrow$; {\em i.e.\/} a magnetic phase with $\langle Z_\uparrow \rangle \neq 0$
and $\langle Z_\downarrow \rangle = 0$ has a finite range of stability.
Thus close to the transition we can neglect the $Z_\downarrow$ field entirely, and transition is in the universality
class of the three-dimensional XY model.

The choice above of $Z_\uparrow$ over $Z_\downarrow$ gives the incorrect
appearance that we are breaking the spin reflection symmetry 
$S_z \rightarrow - S_z$ of $\mathcal{H}$, suggesting the appearance of a net $z$ ferromagnetic moment. 
However, notice that the theory of $Z_\uparrow$ is relativistic,
and so contains both spinons and anti-spinons which carry $S_z = +1/2$ and $S_z = -1/2$ respectively.
The spinon of $Z_\downarrow$ also carries $S_z = -1/2$, and this is degenerate with the anti-spinon
of $Z_\uparrow$ in the O(4) invariant theory in Eq.~(\ref{lhz}). It is this latter degeneracy which is lifted by the DM
interactions, which induce a vector spin chirality along the $z$ direction \cite{kimsenthil} (as we will see below). We will also see
there is no net ferromagnetic moment, because time-reversal symmetry is preserved.

\subsection{Observables}
\label{sec:obs}

To determine the operators corresponding to the ferromagnetic moment, 
let us couple a uniform external field ${\bf h}$ to the lattice Hamiltonian.
This adds to the continuum Lagrangian the term
\begin{equation}
\mathcal{L}_h = - {\bf h} \cdot {\boldsymbol \tau}_{\sigma\sigma'} \left( U^{\ast}_\sigma U_{\sigma'} + V^{\ast}_\sigma V_{\sigma'} +
W^{\ast}_\sigma W_{\sigma'}  \right)
\end{equation}
Inserting the parameterization in Eq.~(\ref{umat}) this becomes
\begin{equation}
\mathcal{L}_h = - {\bf h} \cdot {\boldsymbol \tau}_{\sigma\sigma'} \left( X^{\ast}_\sigma X_{\sigma'} + Y^{\ast}_\sigma Z_{\sigma'} +
Z^{\ast}_\sigma Y_{\sigma'}  \right) \label{lz3}
\end{equation}
We now need to integrate out $X_\sigma$ and $Y_\sigma$ in the Lagrangian $\mathcal{L}_H + \mathcal{L}_{DM} + 
\mathcal{L}_h$ defined by the sum of 
Eqs.~(\ref{lz1}), (\ref{lz2}), and (\ref{lz3}), and collect the terms linear in ${\bf h}$.
Without the DM coupling, we obtain
\begin{equation}
\sim  {\bf h} \cdot {\boldsymbol \tau}_{\sigma \sigma'}
 \left(Z^\ast_\sigma \frac{\partial Z_{\sigma'}}{\partial \tau} - \frac{\partial Z^\ast_\sigma}{\partial \tau} Z_{\sigma'} \right) \label{mg1}
\end{equation}
Comparing with $\mathcal{L}^Z_H$ in Eq.~(\ref{lhz}) we see that this is just the coupling to the conserved SU(2) charges
of the O(4) model: this is the usual term which determines the magnetic susceptibility of the Heisenberg antiferromagnet
\cite{css}. Upon including the effects of $\mathcal{L}_{DM}$ we find that the essential
structure of Eq.~(\ref{mg1}) does not change: the ${\boldsymbol \tau}$ matrices get multiplied by
some $\sigma$-dependent factors ${\boldsymbol \tau}_{\sigma \sigma'} \rightarrow f_\sigma {\boldsymbol \tau}_{\sigma \sigma'}
f_{\sigma'}$ which do not modify the scaling considerations. No term with a new structure
is generated by the DM coupling. It can now be seen that these expressions have vanishing expectation
values under $\mathcal{L}_Z$ in Eq.~(\ref{lz}), and so there is no net ferromagnetic moment in the absence of an external field.

We now turn to the antiferromagnetic order parameter; for a coplanar antiferromagnet, this  is described by
\begin{equation}
{\bf S}_i \propto {\bf N}_1 \cos({\bf Q} \cdot {\bf r}_i ) + {\bf N}_2 \sin({\bf Q} \cdot {\bf r}_i ),
\end{equation}
where ${\bf N}_{1,2}$ are 2 orthogonal vectors representing the spiral order, and ${\bf Q}$ is wavevector
at which the spin structure factor is peaked. For our model, we can see that
\begin{equation}
{\bf N}_1 + i {\bf N}_2 = {\bf S}_u + \zeta {\bf S}_v + \zeta^2 {\bf S}_w . \label{spiral}
\end{equation}
Using Eq.~(\ref{umat}), and keeping only the lowest order term, we therefore obtain \cite{css}
\begin{equation}
{\bf N}_1 + i {\bf N}_2 = \left( \begin{array}{c}
i (Z_\uparrow^2 - Z_\downarrow^2)/2 \\
- (Z_\uparrow^2 + Z_\downarrow^2)/2 \\
-i Z_\uparrow Z_\downarrow
\end{array}
\right);
\end{equation}
in a notation that makes the rotational invariance evident, this relationship is
\begin{equation}
{\bf N}_1 + i {\bf N}_2 = \frac{i}{2} \epsilon_{\alpha\beta} {\boldsymbol \tau}_{\beta\sigma} Z_\sigma Z_\alpha .
\end{equation}
Note that a phase with  $\langle Z_\uparrow \rangle \neq 0$
and $\langle Z_\downarrow \rangle = 0$ has spiral order in the $x$-$y$ plane.

To complete the list of operators which are quadratic in the $Z_\sigma$, we consider the vector spin chirality \cite{kimsenthil}.
This is defined here by
\begin{equation}
{\bf S}_u \times {\bf S}_v + {\bf S}_v \times {\bf S}_w + {\bf S}_w \times {\bf S}_u.
\end{equation}
Using Eq.~(\ref{umat}) we find that the leading operator mapping to vector spin chirality is (dropping an
overall factor of $|Z_\uparrow|^2 + |Z_\downarrow |^2$)
\begin{equation}
Z^\ast_{\sigma} {\boldsymbol \tau}_{\sigma\sigma'} Z_{\sigma'}.
\end{equation}
Note that in the presence of the DM term, the couplings in the effective theory (\ref{lz})
imply that the $z$ component of the vector spin chirality is always non-zero.

\subsection{Critical properties}
\label{sec:crit}

Let us assume the transition to magnetic order proceeds via the condensation of $Z_\uparrow$.
The transition is in the XY universality class, and the dimension of the antiferromagnetic order parameter is
\begin{equation}
\mbox{dim}[{\bf N}_1 ] = \mbox{dim}[{\bf N}_2] = \mbox{dim}[Z_\uparrow^2]  = \frac{1 + \overline{\eta}}{2}\;.
\end{equation}
The value of the exponent $\overline{\eta}$ can be read off from results for the three-dimensional
XY model \cite{vicari3,vicari4}, and we obtain $\overline{\eta} \approx 1.474$.
The antiferromagnetic susceptibility will diverge at the critical point as $T^{-(2-\overline{\eta})}$.
We note the recent work of Ref.~\onlinecite{tarun} in a different context, which also considered
a model with an XY critical point at which the physically measurable magnetic order was
the square of the XY field. 

The behavior of the uniform magnetic susceptibility follows from the scaling dimension of the
operators in Eq.~(\ref{mg1}). For ${\bf h}$ along the $z$ direction, the magnetization is the just
the conserved U(1) charge of the XY model: so \cite{cs} it has scaling dimension 2, and the susceptibility
$\sim T$. For ${\bf h}$ along the $x$ or $y$ directions, we have to integrate out $Z_\downarrow$,
and then the lowest dimension operator coupling to the square of the field is $|Z_\uparrow|^2$.
This means that the susceptibility only has a weak singularity at the quantum critical point 
given by that in $\langle |Z_\uparrow |^2 \rangle$: at the quantum critical point, there is a non-analytic
term $\sim T^{3-1/\nu}$, above an analytic background.

\section{Conclusion}
\label{sec:conclusion}

We have presented a theory for the quantum critical point between a $Z_2$ spin liquid
and an ordered antiferromagnet for the kagom\'e antiferromagnet in the presence of DM interactions.
The critical theory is just the three dimensional XY model. However, the XY order parameter
carries a $Z_2$ gauge charge, and so it is not directly observable. In particular, the antiferromagnetic
order parameter is the square of the XY order parameter. Specifically, the theory is given by
$\mathcal{L}_Z$ in Eq.~(\ref{lz}), and its observables are described in Section~\ref{sec:obs}.

It is interesting to compare our results with recent observations of quantum critical scaling
in ZnCu$_3$(OH)$_6$Cl$_2$ by Helton
{\em et al.} \cite{helton2}. Their neutron scattering measurements show an antiferromagnetic
susceptibility which scales as $T^{-0.66}$. This is actually in reasonable agreement with our
theory, which has a susceptibility $ \sim T^{-0.526}$. However, they also observe a similar divergence in 
measurements of the uniform magnetization, while our theory only predicts a very weak singularity.
We suspect that this difference is due to the present of impurities \cite{bert,gregor,chitra,ka5,mila2,rajiv}, which
can mix the uniform and staggered components. A complete study of impurities near the quantum critical 
point described above is clearly called for.

\acknowledgements

We thank L.~Messio, O.~C\'epas, C.~Lhuillier, and E.~Vicari for useful discussions.
This research was supported by the National Science Foundation under grant DMR-0757145, by the FQXi
foundation, and by a MURI grant from AFOSR. Y.H. is also supported in part by a Samsung scholarship.

\appendix
\section{Microscopic form of the Hamiltonian}
\label{sec:micro}
We here give explicit expressions for the Hamiltonian introduced in Sec.~\ref{sec:model}. We introduce the following set of unit vectors
\begin{eqnarray}
{\bf{e}}_1&=& a\left(\frac{1}{2},\frac{\sqrt{3}}{2}\right)\nonumber \\ {\bf{e}}_2&=&a\left(\frac{1}{2},-\frac{\sqrt{3}}{2}\right) \nonumber \\ {\bf{e}}_3&=&  a(-1,0)
\label{eq:direct}
\end{eqnarray}
which allows to express $k_i = {\bf{k}} \cdot {\bf{e}}_i$. In the following we concentrate on the states with $3$ sites per unit cell. In that case the vector $\Psi$ introduced in Eq.~\eqref{eq:Psi} has $6$ components and the matrix $\mathbb{C}$ consequently is a $3\times3$ matrix with the following entries:
\begin{widetext}
\begin{eqnarray}
\mathbb{C}_{uv}&=& \frac{J}{2} q_1^* e^{ik_1} + \frac{J}{2} q_2^* e^{-ik_1} + \frac{iD}{4}p_1^* e^{ik_1} + \frac{iD}{4}q_1^* e^{ik_1} + \frac{iD}{4}p_2^* e^{-ik_1} + \frac{iD}{4}q_2^* e^{-ik_1}  \nonumber \\
\mathbb{C}_{uw}&=& -\frac{J}{2} q_1^* e^{-ik_3} - \frac{J}{2} q_2^* e^{ik_3} - \frac{iD}{4}p_1^* e^{-ik_3} + \frac{iD}{4}q_1^* e^{-ik_3} - \frac{iD}{4}p_2^* e^{ik_3} + \frac{iD}{4}q_2^* e^{ik_3}  \nonumber \\
\mathbb{C}_{vw}&=& \frac{J}{2} q_1^* e^{ik_2} + \frac{J}{2} q_2^* e^{-ik_2} + \frac{iD}{4}p_1^* e^{ik_2} + \frac{iD}{4}q_1^* e^{ik_2} + \frac{iD}{4}p_2^* e^{-ik_2} + \frac{iD}{4}q_2^* e^{-ik_2}  \nonumber \\
\mathbb{C}_{vu}&=& -\frac{J}{2} q_1^* e^{-ik_1} - \frac{J}{2} q_2^* e^{ik_1} - \frac{iD}{4}p_1^* e^{-ik_1} + \frac{iD}{4}q_1^* e^{-ik_1} - \frac{iD}{4}p_2^* e^{ik_1} + \frac{iD}{4}q_2^* e^{ik_1}  \nonumber \\
\mathbb{C}_{wv}&=& -\frac{J}{2} q_1^* e^{-ik_2} - \frac{J}{2} q_2^* e^{ik_2} - \frac{iD}{4}p_1^* e^{-ik_2} + \frac{iD}{4}q_1^* e^{-ik_2} - \frac{iD}{4}p_2^* e^{ik_2} + \frac{iD}{4}q_2^* e^{ik_2}  \nonumber \\
\mathbb{C}_{wu}&=&  \frac{J}{2} q_1^* e^{ik_3} + \frac{J}{2} q_2^* e^{-ik_3} + \frac{iD}{4}p_1^* e^{ik_3} + \frac{iD}{4}q_1^* e^{ik_3} + \frac{iD}{4}p_2^* e^{-ik_3} + \frac{iD}{4}q_2^* e^{-ik_3}\;.
\label{eq:cmat}
\end{eqnarray}
\end{widetext}

\section{Dispersion of the lowest excited spinon state}
\begin{figure}[h]
\includegraphics[width=0.5\textwidth]{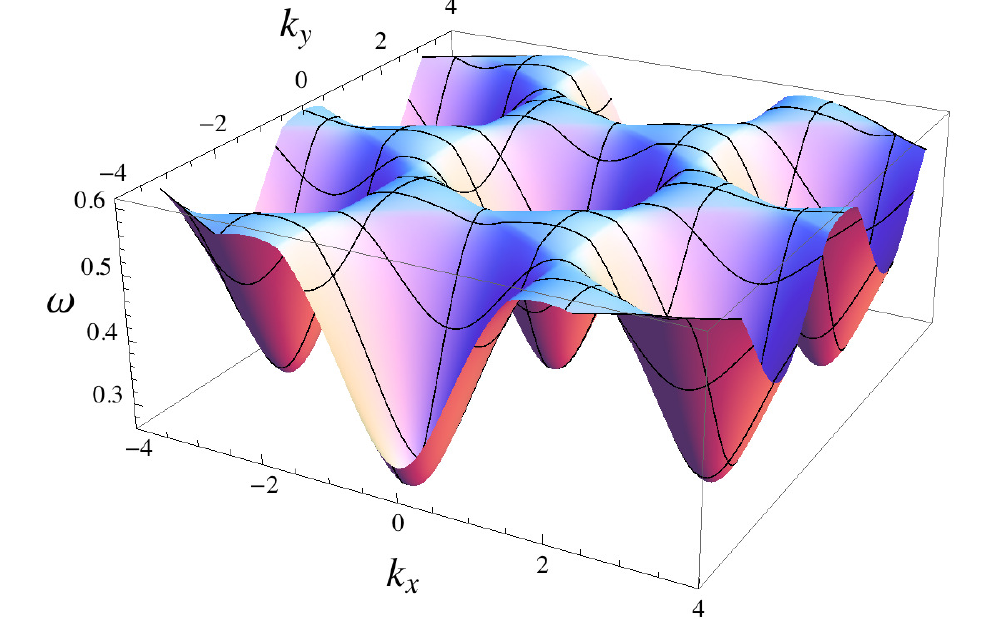}
\caption{Momentum dependence of the energy $\omega(k)$ of the lowest excited spinon state of the kagome-lattice quantum antiferromagnet for the $q_1=q_2$ state at $D/J=0.3, \kappa=0.2$. The minimum excitation energy is at $\mathbf{k}=0$ and has a finite energy gap.}
\label{fig:D3K2pos}
\end{figure}

\begin{figure}[h]
\includegraphics[width=0.5\textwidth]{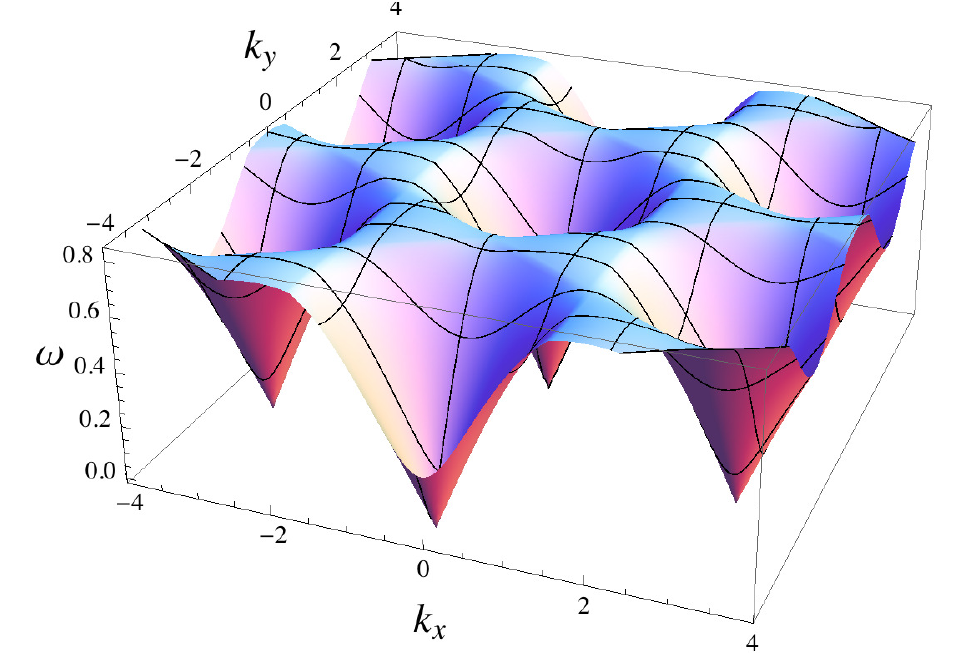}
\caption{Dispersion of the lowest excited spinon state for the $q_1=q_2$ state at $D/J=0.3, \kappa=0.4$. The energy gap closes at $\mathbf{k}=0$ and condensation occurs. }
\label{fig:D3K4pos}
\end{figure}

For $q_1=q_2$ and $D=0$, the ground state is doubly degenerate. The degeneracy splits as $D$ moves away from 0. In all cases, the minimum excitation energy occurs at $\mathbf{k}=0$. \\
Fig.~\ref{fig:D3K2pos} plots the momentum dependence of the energy of the lowest excited spinon at $D/J=0.3, \kappa=0.2$. There is a finite energy gap at $\mathbf{k}=0$ making it a gapped spin liquid. On the other hand, Fig.~\ref{fig:D3K4pos} is the dispersion plot for a case with long-range magnetic order. The energy gap at $\mathbf{k}=0$ is closed and condensation occurs at this wave vector. 

\begin{figure}[h]
\includegraphics[width=0.5\textwidth]{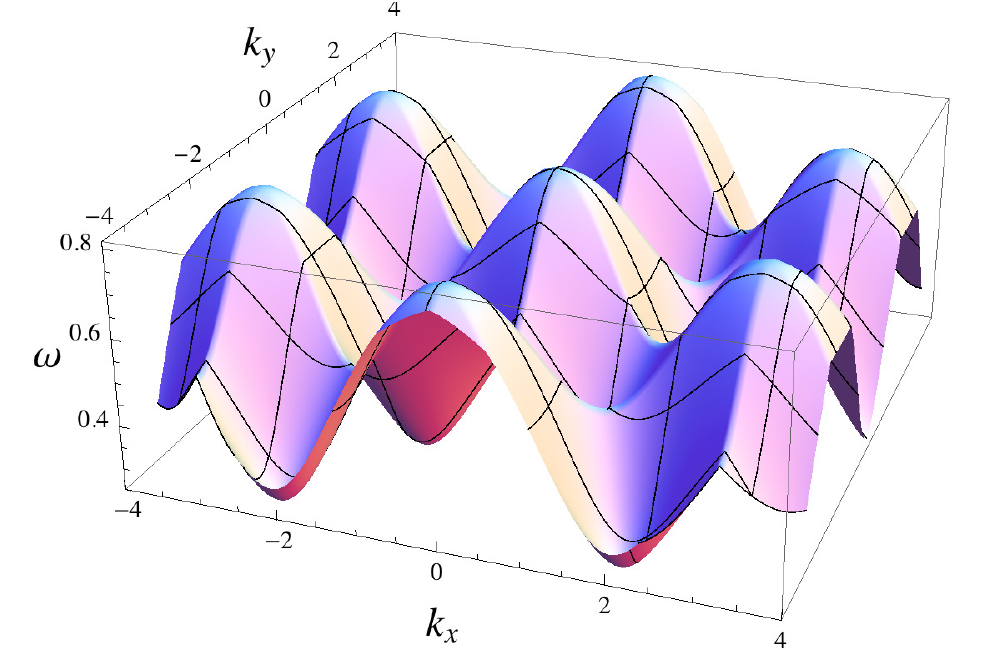}
\caption{disorder: $q_1=-q_2, D/J=0.05, \kappa=0.3$}
\label{fig:D05K3neg}
\end{figure}

\begin{figure}[h]
\includegraphics[width=0.5\textwidth]{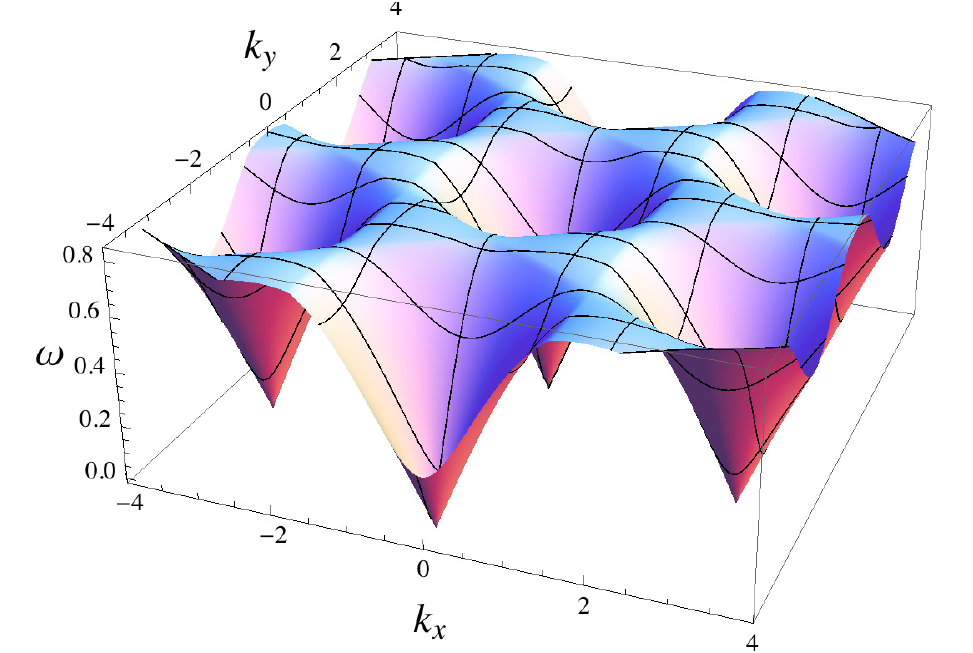}
\caption{order: $q_1=-q_2, D/J=0.03, \kappa=0.55$}
\label{fig:D03K55neg}
\end{figure}

For $q_1=-q_2$, there is a unique ground state even for $D=0$. Energy minima is at $\mathbf{k}=\pm(2\pi/3,0)$. Fig. ~\ref{fig:D05K3neg} shows the dispersion of the lowest lying state of the spin liquid with $D/J=0.03, \kappa=0.55$. A case with long range ordering is shown in Fig. ~\ref{fig:D03K55neg}.    

\section{Condensation}
For the $q_1=q_2$ state Eq.~\eqref{umat} leads to the following parametrization of the condensation of $Z_\downarrow$ field at $\mathbf{k}=0$.
\begin{eqnarray}
\left( \begin{array} {c} x_u^{\uparrow} \\ x_u^{\downarrow} \end{array} \right) &=& \ell \left( \begin{array} {c} i  \\ 1 \end{array} \right) \nonumber \\
\left( \begin{array} {c} x_v^{\uparrow} \\ x_v^{\downarrow} \end{array} \right) &=& \ell \left( \begin{array} {c} i\z^2  \\ \z \end{array} \right)\nonumber \\
\left( \begin{array} {c} x_w^{\uparrow} \\ x_w^{\downarrow} \end{array} \right) &=& \ell \left( \begin{array} {c} i\z  \\ \z^2 \end{array} \right)
\end{eqnarray}
where $\ell$ is the size of the condensate. 
Condensation of $Z_\uparrow$ can be written similarly. For $D>0$, $Z_\downarrow$ field condensation is energetically favored while the opposite is true for $D<0$. The two condensations are degenerate for $D=0$.

For the $q_1=-q_2$ state, condensation occurs at $\tilde{k}\equiv\vec{k}=(2\pi/3,0)$ or $\vec{k}=-\tilde{k}$. The two states have identical energies and the condensation is spontaneously chosen. Similar analysis to Eqns.~\eqref{eq:lagra} and~\eqref{umat} gives the eigenvectors corresponding to the lowest lying state. For $\vec{k}=\tilde{k}$ this is $(i,-i,i,1,-1,1)$ while for $\vec{k}=-\tilde{k}$ the corresponding eigenvector is $(-i,i,-i,1,-1,1)$. 
Therefore condensations can be parametrized as  
\bea
\left(\begin{array} {c} x^u_{k\uparrow} \\ x^u_{-k\downarrow} \end{array}  \right)&=&\ell \left( \begin{array}{c} i \\ 1 \end{array}\right) \nn\\
\left(\begin{array} {c} x^v_{k\uparrow} \\ x^v_{-k\downarrow} \end{array}  \right)&=&\ell \left( \begin{array}{c} -i \\ -1 \end{array}\right) \nn\\
\left(\begin{array} {c} x^w_{k\uparrow} \\ x^w_{-k\downarrow} \end{array}  \right)&=&\ell \left( \begin{array}{c} i \\ 1 \end{array}\right) \nn\\
\eea
for  $\vec{k}=\tilde{k}$, and 
\bea
\left(\begin{array} {c} x^u_{k\uparrow} \\ x^u_{-k\downarrow} \end{array}  \right)&=&\ell \left( \begin{array}{c} -i \\ 1 \end{array}\right) \nn\\
\left(\begin{array} {c} x^v_{k\uparrow} \\ x^v_{-k\downarrow} \end{array}  \right)&=&\ell \left( \begin{array}{c} i \\ -1 \end{array}\right) \nn\\
\left(\begin{array} {c} x^w_{k\uparrow} \\ x^w_{-k\downarrow} \end{array}  \right)&=&\ell \left( \begin{array}{c} -i \\ 1 \end{array}\right) \nn\\
\eea
for  $\vec{k}=-\tilde{k}$.

\end{document}